\documentclass{aa}

\usepackage{graphicx}
\usepackage{txfonts}
\usepackage{lipsum}
\usepackage{natbib}
\usepackage{subcaption}         
\usepackage{lscape}             
\usepackage{placeins}           

\begin{document}

   \title{Local luminosity functions of galaxies at (sub)millimeter wavelengths from \textit{Planck} surveys}
   \titlerunning{Local luminosity functions at (sub)millimeter wavelengths}

%

   \author{Matteo Bonato\inst{1}
          \and
          Gianfranco De Zotti\inst{2}
          }
\institute{INAF, Istituto di Radioastronomia - Italian ARC, Via Piero Gobetti 101, I-40129 Bologna, Italy\\
\email{matteo.bonato@inaf.it}
\and
INAF, Osservatorio Astronomico di Padova, Vicolo Osservatorio 5, I-35122 Padova, Italy\\
\email{gianfranco.dezotti@inaf.it}}
   \date{Received ...}


  \abstract
  {} 
   {The \textit{Planck} all-sky surveys at (sub)millimeter wavelengths enable us to accurately determine the corresponding local luminosity functions up to the highest luminosities. The detected galaxies are strictly local, so evolutionary effects, which are known to be particularly strong at these wavelengths, are not a problem. However, previous studies have so far relied only on the \textit{Planck} Early Release Compact Source Catalog (ERCSC) for this purpose. Another important improvement over earlier estimates is the availability of complete all-sky catalogs of galaxies within hundreds of megaparsecs, with redshift-independent distances for nearby objects, for which redshifts are not reliable distance estimators. In this paper we re-estimate the (sub)millimeter local luminosity functions using data from the Second \textit{Planck} Catalog of Compact Sources, which supersedes the previously used ERCSC and contains far more sources and more accurate photometry. }
   {We computed the luminosity functions using both the classical $1/V_{\rm max}$ and the nonparametric kernel density estimation (KDE) method, which overcomes limitations of binning techniques. Our implementation of the KDE uses the $1/V_{\max}$ weighting to account for survey selection effects.}
   {We obtain \textit{Planck}-based local luminosity functions at 857, 545,  353, and 217 GHz, as well as the total IR luminosity function and the dust mass function. We find significant differences from earlier estimates and discuss their possible origins.}
   {}


   \maketitle

\section{Introduction}
The local luminosity function (LLF) is the necessary benchmark for assessing the evolutionary properties of extragalactic sources. The first estimates of the galaxy LLF at submillimeter (submm) wavelengths based on blind surveys were obtained with the Balloon-borne Large Aperture Submillimeter Telescope (BLAST), which surveyed $\simeq 8.7\,\hbox{deg}^2$ at 250, 350, and $500\,\mu$m \citep{Eales2009}. Surveys with the Spectral and Photometric Imaging REceiver (SPIRE) and the Photodetector Array Camera and Spectrometer (PACS) on board the \textit{Herschel} Space Observatory provided better statistics and improved angular resolution, enabling firmer identifications  \citep{Eales2010, Oliver2012, Lutz2014}; areas of a few tens of square degrees were used for luminosity function (LF) estimates \citep{Vaccari2010, Gruppioni2013, Marchetti2016}.

The shallow \textit{Planck} all-sky surveys with the High Frequency Instrument (HFI) are ideal for sampling the LLF up to the highest luminosities, without being affected by the strong evolutionary effects present at submm wavelengths. They are the only blind surveys that allow estimates of the galaxy LLF at 850 and 1382$\,\mu$m, since no other instrument has covered a sufficiently large area at these wavelengths. So far, only a preliminary exploitation of \textit{Planck}/HFI data has been published. \citet{Negrello2013} used the \textit{Planck} Early Release Compact Source Catalog \citep[ERCSC;][]{PlanckERCSC2011} to derive the LLF of dusty galaxies in the local Universe (distances $\la 100\,$Mpc) at 350, 500, and $850\,\mu$m and also to provide a preliminary estimate at 1.382\,mm.

The \textit{Planck} galaxy catalogs have been subsequently much improved \citep{PCCS2014, PCCS22016, PlanckBeeP2020}. While the ERCSC used data from one complete survey and 60\% of the second survey, the Second \textit{Planck} Catalog of Compact Sources (PCCS2) lists discrete objects extracted from the full mission maps. These include five surveys with the HFI and eight surveys with the Low Frequency Instrument (LFI). The PCCS2 not only reaches significantly fainter flux densities but also takes advantage of a much better knowledge of the instrument, leading to improved calibration and map quality, as well as better beam characterization, which is crucial for obtaining precise flux density measurements. Despite the extensive work carried out by \citet{Negrello2013} to account for various problems affecting ERCSC flux densities, substantial uncertainties remained.

Catalogs were produced for each of the nine \textit{Planck} frequency channels. The HFI catalogs (at 100, 143, 217, 353, 545, and 857\,GHz, corresponding to 3000, 2100, 1382, 850, 550, and $350\,\mu$m, respectively) were divided into two subcatalogs, PCCS2 and PCCS2E, based on source reliability, defined as the fraction of sources above a given signal-to-noise ratio (S/N)  that are real. The PCCS2 catalog contains sources whose reliability could be quantified and was found to be at least 80\%, while PCCS2E contains sources of unknown reliability. The S/N threshold corresponding to the 80\% reliability is generally in the range 4--5 in the high Galactic latitude area of interest here but may increase to up to 10 around the Galactic plane. The right-hand panels of Fig.\,8 of \citet{PCCS22016} show maps of such S/N thresholds for all HFI channels.

\citet{PlanckBeeP2020} present a reanalysis of all PCCS2 and PCCS2E (PCCS2$+$2E) sources detected at 857 GHz (Bayesian Extraction and Estimation Package (BeeP) catalog). In this paper, data re-extracted from \textit{Planck} and Infrared Astronomical Satellite (IRAS) maps are combined to assess the reliability of each source and to redetermine the flux densities at 857, 545, 353, and 217\,GHz. It also contains flux densities at 3000\,GHz ($100\,\mu$m) derived from the corresponding IRIS map \citep[a reprocessed IRAS map;][]{MivilleDeschenesLagache2005}, using the same pixelization as the \textit{Planck} maps.

Singling out dusty galaxies in the ERCSC was another laborious task \citep{Negrello2013}. In recent years, the situation has greatly improved in this respect. \citet{Kovlakas2021} compiled an all-sky catalog of 204,733 galaxies within a distance of 200\,Mpc, called the Heraklion Extragalactic Catalog (HECATE). This catalog contains accurate positions, galaxy sizes, distances (redshift-independent as far as possible for the nearest objects), multiband photometry, and several global properties of galaxies.

The source of $z$-independent distances in HECATE is the NASA/IPAC Extragalactic Database (NED) master list of redshift-independent extragalactic distances (NED-D\footnote{\url{https://ned.ipac.caltech.edu/Library/Distances/}}). The NED-D catalog was compiled through an extensive literature search of distances based on primary methods (standard candles or standard rulers) or secondary methods, such as the Tully-Fisher or fundamental plane relations. The methodology is described in \citet{Steer2017}. For galaxies not in NED-D, \citet{Kovlakas2021} used spectroscopic redshifts. To avoid the effects of peculiar velocities and local deviations from the Hubble flow, they used $z$-independent distances as the training data set to infer distances at similar recession velocities. The HECATE catalog was estimated to be complete for $\log(L_b/L_\odot)\ga 10$ up to distances of $\simeq 100\,$Mpc.

The growth of large-scale imaging surveys in different wavebands and the constantly increasing completeness of redshift and redshift-independent distance measurements in the regularly updated NED have allowed the construction of the NED Local Volume Sample \citep[NED-LVS;][]{Cook2023}. The NED-LVS catalog contains $\sim 1.9\times 10^6$ objects with distances of up to 1000\,Mpc. It was found to be nearly 100\% complete out to $\sim 400\,$Mpc for galaxies with $L_{Ks}\ge L_\star$ and about 80\% complete out to 180\,Mpc with no luminosity cut. Redshift-independent distances, when available, were chosen at distances less than 200\,Mpc. The NED-LVS catalog was found to be 10--20\% more complete than HECATE and is adopted in the following. These catalogs of local galaxies allow us to identify \textit{Planck}-detected galaxies by cross-matching \textit{Planck} catalogs with them.

In this paper we take advantage of the much improved data situation to obtain novel determinations of the LLF at 857, 545, 353, and 217\,GHz. We describe the data in Sect.\,\ref{Sect_data} and present the two methods adopted to compute the LLFs in Sect.\,\ref{Sect_methodology} . We also used the 857\,GHz sample to derive the total infrared (IR; 8--$1000\,\mu$m) LF because it corresponds to the \textit{Planck} frequency closest to the dust emission peak. Following a widely used approach \citep{Dunne2000, Vlahakis2005, Clemens2013}, we estimate the dust mass function based on the 353\,GHz sample. Section~\ref{sect:conclusions} discusses the results and summarizes our conclusions.

Throughout this paper we adopted a spatially flat cosmology with $H_0=67.7\,\hbox{km}\,\hbox{s}^{-1}\,\hbox{Mpc}^{-1}$ and $\Omega_m=0.31$ \citep{PlanckParameters2020}. Whenever the NED-LVS distances were derived from redshift, we adopted a Hubble constant $H_0=69.6\,\hbox{km}\,\hbox{s}^{-1}\,\hbox{Mpc}^{-1}$ and a flat cosmology with $\Omega_m=0.3$. We applied a small correction for the different cosmological model used here.

\section{Data}\label{Sect_data}

This paper is based on the PCCS2$+$2E catalogs \citep{PCCS22016} at 217, 353, 545, and 857\,GHz, downloaded from the \textit{Planck} Legacy Archive (PLA\footnote{\url{https://pla.esac.esa.int/pla/#catalogues}}). These remain the most up-to-date \textit{Planck} products.

We restricted our analysis to galaxies at Galactic latitudes $|b|\ge 30^\circ$ to minimize the contamination of the flux densities by diffuse Galactic emission. This region almost entirely comprises the ``extragalactic zone,'' which, for HFI channels, is defined as the region where the source reliability can be accurately quantified. For each HFI frequency channel the PLA provides a ``zone mask'' where source reliability is unquantified; the ``extragalactic zone'' is thus the region outside the ``zone mask'' (see Fig. 8 of \citealt{PCCS22016} for an example).

The PCCS2 provides four flux density measures for each source: DETFLUX, supplied by the source detection algorithm; aperture photometry (APERFLUX); fitting a model of the point spread function at the position of the source (PSFFLUX); and fitting an elliptical Gaussian model to the source (GAUFLUX). We used APERFLUX for all channels except 857\,GHz, where we used BeeP photometry, because \citet{PCCS22016} find APERFLUX to be the most appropriate flux-density measure to use for the higher frequency channels.

The BeeP catalog contains exactly the same sources as the original PCCS2$+$2E 857\,GHz catalog. \citet{PlanckBeeP2020} did not attempt to detect new sources, but remeasured the photometric data of the cataloged sources. Thus, we relied on the PCCS2$+$2E to determine the completeness limits. In the extragalactic zone they are 791\,mJy for the PCCS2 and 927\,mJy for the PCCS2E at the 90\% level.

The 857\,GHz selection implies that the BeeP catalog is not suitable for defining complete samples at the other \textit{Planck} frequencies. We therefore referred directly to the PCCS2$+$2E catalogs.

\section{Methodology}\label{Sect_methodology}

We computed the LFs using the classical $1/V_{\rm max}$ \citep{Schmidt1968} and the kernel density estimation (KDE) method \citep{CaditzPetrosian1993, Yuan2020, Yuan2022}.

The $1/V_{\rm max}$ method is a nonparametric technique that starts with dividing the data into luminosity bins. Each source is weighted by the reciprocal of the maximum volume, $V_{\rm max}$, within which it can be detected. We confined our analysis to distances $D<180\,$Mpc, at which NED-LVS is expected to be complete for the relatively bright galaxies considered here \citep{Cook2023}. For these nearby galaxies, $V_{\rm max}$ is given by
\begin{equation}
    \label{eq:Vmax}
 V_{\rm max}=\frac{\omega}{3}D_{\rm max}^3,
   \end{equation}
where $\omega=2\pi$ is the solid angle covered by the sample and $D_{\rm max}=\min(D_{\rm L,max}, D_{\rm lim})$, with $D_{\rm lim}=180\,$Mpc and $D_{\rm L,max}=(L\,K(z)/4\,\pi\,S_{\rm lim})^{1/2}$ is the distance at which the source would be at the detection limit, $S_{\rm lim}$. The K-correction, $K(z)$, is not completely negligible for the most distant objects because of the steepness of the dust emission spectrum at (sub)mm wavelengths. We computed it using the spectral energy distribution (SED) of starburst galaxies defined by \citet{Cai2013}, which is available online\footnote{ \url{http://staff.ustc.edu.cn/~zcai/galaxy_agn/index.html}}. 

The space density in the i-th luminosity bin, with central luminosity $L_i$ is
   \begin{equation}\label{eq:LFun}
      \Phi(L_i)=\sum_{j=1}^{N_{{\rm bin},i}}\frac{1}{V_{{\rm max},i}},
   \end{equation}
where the sum is over the $N_{{\rm bin},i}$ sources in the bin.

We computed the errors following \citet[][their Appendix A]{Ananna2022}.
The upper and lower limits of the 0.8413 confidence level to $\Phi(L_i)$ are given by
\begin{equation}\label{eq:LFun_err}
\begin{aligned}
\Phi_{\rm up}(L_i)=& W_{\mathrm{eff}, i} \times \kappa_{\rm up}(N_{\mathrm{eff}, i})\\
\Phi_{\rm low}(L_i)=& W_{\mathrm{eff}, i} \times \kappa_{\rm low}(N_{\mathrm{eff}, i}).
\end{aligned}
\end{equation}
The weight $W_{\rm eff}$ and the effective number of sources in the bin, $N_{\rm eff}$, are given by
\begin{equation}\label{eq:Neff}
\begin{aligned}
W_{\mathrm{eff}, i} =& \left(\sum_{j=1}^{N_{{\rm bin},i}} \frac{1}{V_{\mathrm{max}, j}^2}\right) \times \left(\sum_{j=1}^{N_{{\rm bin},i}} \frac{1}{V_{\mathrm{max}, j}}\right)^{-1}\\
N_{\mathrm{eff}, i} =& \left(\sum_{j=1}^{N_{{\rm bin},i}} \frac{1}{V_{\mathrm{max}, i}} \right) \times \left(W_{\mathrm{eff}, i}\right)^{-1}.
\end{aligned}
\end{equation}
We computed the functions $\kappa_{\rm up}$ and $\kappa_{\rm low}$ using eqs. (9) and (14) of \citealt{Gehrels1986}:
\begin{equation}\label{eq:Gehrels}
\begin{aligned}
\kappa_{\rm up}=& (1+N_{\mathrm{eff}})\left[1-\frac{1}{9(1+N_{\mathrm{eff}})}+\frac{1}{3(1+N_{\mathrm{eff}})^{1/2}}\right]^3\\
\kappa_{\rm low}=& N_{\mathrm{eff}}\left(1-\frac{1}{9 N_{\mathrm{eff}}}-\frac{1}{3 N_{\mathrm{eff}}^{1/2}}\right)^3.
\end{aligned}
\end{equation}

The KDE is a nonparametric approach for estimating continuous density functions that avoids the discretization artifacts inherent to traditional binning techniques while properly accounting for survey selection effects. We followed the adaptive framework described in \citet{Yuan2022}, accounting for survey selection effects through $1/V_{\max}$ weighting.

The weighted KDE estimator for the LF $\Phi(\mathcal{L})$, with $\mathcal{L} = \log_{10}L$, is defined as
\begin{equation}
    \Phi(\mathcal{L}) = \sum_{i=1}^{N} \frac{1}{V_{\text{max}, i}} \frac{1}{\sqrt{2\pi}h} \exp \left[ -\frac{(\mathcal{L} - \mathcal{L}_i)^2}{2h^2} \right] \,,
\end{equation}
where $\mathcal{L}_i$ are the observed log-luminosities, $h$ is the smoothing bandwidth, and the sum is over the $N$ galaxies in the sample. This formulation ensures that each galaxy contributes to the total density estimate in proportion to its accessible volume, providing a continuous and smooth representation of the LF.

A critical aspect of KDE is the choice of bandwidth, which governs the smoothing scale. Too large a bandwidth can lead to over-smoothing; in the case of a sharp steepening of the LF, this would generate spurious excesses by scattering sources from below to above the knee. Conversely, too narrow a bandwidth results in a noisy estimator dominated by Poisson fluctuations. While automated selection criteria exist, such as the widely used ``rule of thumb'' methods by \citet{Silverman1986} and \citet{Scott1992}, these are often optimized for Gaussian-like distributions and may lead to significant over-smoothing when applied to functions with a sharp break or an exponential tail, such as the LF. We therefore performed an empirical optimization by testing a range of bandwidths and comparing the resulting KDEs with the binned $1/V_{\rm max}$ estimates.
We find that a constant bandwidth of 0.25 (in log-luminosity space) provides the optimal balance between bias and variance. This value ensures that the steep decline at the bright end of the LF is accurately preserved without introducing spurious oscillations driven by local fluctuations.

We estimated statistical uncertainties via bootstrap resampling. We generated 200 realizations of the source catalog and adopted the median of the resulting distributions as our best estimate, with the 16th and 84th percentiles defining the $1\sigma$ confidence interval.

\section{Luminosity functions}\label{SectLF}

\begin{figure}
   \resizebox{\hsize}{!}
            {\includegraphics {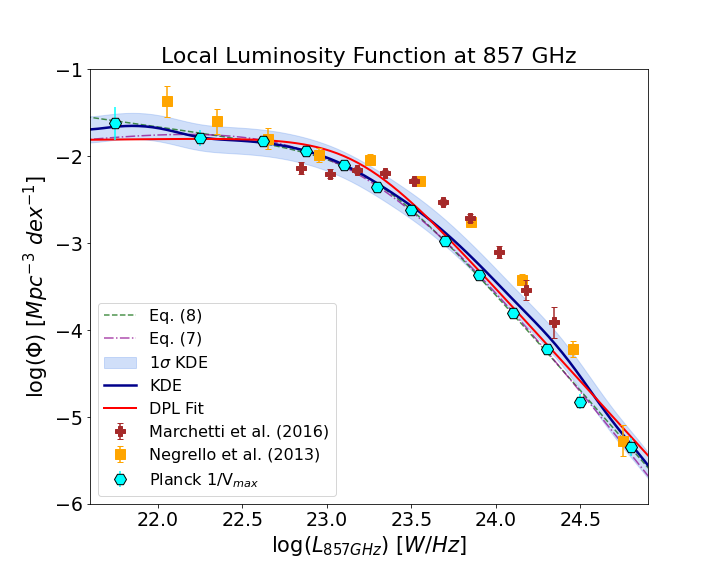}}
      \caption{Local luminosity function at 857\,GHz determined using both the $1/V_{\rm max}$ and KDE estimators. In the $1/V_{\rm max}$ case, we also show the double power-law fits with eqs.\,(\ref{eq:Lawrence}) and (\ref{eq:Massardi}). In the KDE case, the solid dark blue line shows the central estimates, while the red line (DPL fit) shows the double power-law fit (eq.\,(\ref{eq:Massardi})). Earlier estimates by \citet{Negrello2013} and \citet{Marchetti2016} are also shown for comparison. }
         \label{LF857}
\end{figure}

\begin{table}
\caption{Local luminosity function of galaxies at 857\,GHz obtained with the $1/V_{\rm max}$ method.}               
\label{table:857LLF}    
\centering
\setlength{\tabcolsep}{5pt}                        
\begin{tabular}{c r r c r r}
\hline\hline               
\multicolumn{1}{c}{$\log(L_{857\,\rm GHz}$)} & \multicolumn{1}{c}{$N$} & \multicolumn{1}{c}{$N_{\rm eff}$} & \multicolumn{1}{c}{$\log(\Phi)$}  & \multicolumn{1}{c}{err$+$} & \multicolumn{1}{c}{err$-$} \\
\multicolumn{1}{c}{$\hbox{W}\,\hbox{Hz}^{-1}$} & & & \multicolumn{1}{c}{$\hbox{Mpc}^{-3}\,\hbox{dex}^{-1}$} \\
\hline                      
    21.750   &      9  &   6.749  &  -1.622  &   0.191  &   0.204  \\
    22.250   &     34  &  26.780  &  -1.777  &   0.091  &   0.093  \\
    22.625   &     54  &  45.255  &  -1.828  &   0.065  &   0.066  \\
    22.875   &     96  &  89.878  &  -1.945  &   0.048  &   0.048  \\
    23.100   &    115  & 110.096  &  -2.106  &   0.043  &   0.043  \\
    23.300   &    130  & 124.289  &  -2.357  &   0.041  &   0.041  \\
    23.500   &    145  & 139.443  &  -2.624  &   0.038  &   0.038  \\
    23.700   &    129  & 123.535  &  -2.977  &   0.041  &   0.041  \\
    23.900   &    106  & 102.330  &  -3.371  &   0.045  &   0.045  \\
    24.100   &     85  &  81.426  &  -3.804  &   0.051  &   0.051  \\
    24.300   &     63  &  60.373  &  -4.223  &   0.059  &   0.060  \\
    24.500   &     30  &  29.522  &  -4.824  &   0.086  &   0.088  \\
    24.800   &     22  &  22.000  &  -5.347  &   0.101  &   0.103  \\
\hline                                  
\end{tabular}
\tablefoot{The first column contains the bin center; the second, the number of sources; the third, the effective number given by eq.\,(A2) of \citet{Ananna2022}; the fourth, the space density; and the last two, the positive and negative Poisson errors on $\log(\Phi)$, computed following \citet{Gehrels1986}.}
\end{table}

\begin{table}
    \centering
    \caption{Best-fit parameters for the double power-law fits to the LLFs at 857, 545, 353, and 217\,GHz.}
    \label{table:LFfit}
    \renewcommand{\arraystretch}{1}
    \begin{tabular}{lccrr}
    \hline
    \hline
Model & \multicolumn{1}{c}{$\log_{10}(\Phi_\star)$} & \multicolumn{1}{c}{$\log_{10}(L_\star)$} & \multicolumn{1}{c}{$\alpha$} & \multicolumn{1}{c}{$\beta$} \\
  & \multicolumn{1}{c}{$\hbox{Mpc}^{-3}\,\hbox{dex}^{-1}$} & \multicolumn{1}{c}{$\hbox{W}\,\hbox{Hz}^{-1}$} & &  \\
    \hline
    \hline
    \multicolumn{5}{c}{\textit{857\,\rm GHz}} \\
    \hline
    Eq.\,(\ref{eq:Lawrence}) & -1.506 & 22.763 &  0.765 & 2.734 \\
    Eq.\,(\ref{eq:Massardi}) & -1.953 & 22.850 &  0.264 & 2.649 \\
    KDE                      & -1.779 & 23.186 & -0.021 & 2.137 \\
    \hline
    \hline
    \multicolumn{5}{c}{\textit{545\,\rm GHz}} \\
    \hline
    Eq.\,(\ref{eq:Lawrence}) & -1.512 & 22.457 &  0.992 & 3.374 \\
    Eq.\,(\ref{eq:Massardi}) & -1.963 & 22.850 &  0.264 & 2.649 \\
    KDE                      & -2.029 & 23.007 &  0.176 & 3.012 \\
    \hline
    \hline
    \multicolumn{5}{c}{\textit{353\,\rm GHz}} \\
    \hline
    Eq.\,(\ref{eq:Lawrence}) & -1.400 & 21.864 & 0.790 & 3.437 \\
    Eq.\,(\ref{eq:Massardi}) & -1.854 & 22.336 &  0.093 & 2.705 \\
    KDE                      & -2.001 & 22.449 &  0.202 & 2.600 \\
    \hline
    \hline
    \multicolumn{5}{c}{\textit{217\,\rm GHz}} \\
    \hline
    Eq.\,(\ref{eq:Lawrence}) & -1.668 & 21.635 & 1.313  & 3.230 \\
    Eq.\,(\ref{eq:Massardi}) & -2.063 & 21.986 & 0.270  & 4.194 \\
    KDE                      & -2.372 & 22.137 & 0.446 & 3.388 \\
    \hline
    \hline
    \end{tabular}
\tablefoot{In the $1/V_{\rm max}$ case we fitted both eq.\,(\ref{eq:Lawrence}) and eq.\,(\ref{eq:Massardi}); in the KDE case we fitted only eq.\,(\ref{eq:Massardi}).}
\end{table}

\subsection{The 857\,GHz ($350\,\mu$m) LLF}

There are 1361 sources at $|b|\ge 30^\circ$ with $\hbox{S/N} \ge 5$ at 857 GHz. Only 28 of them have $S_{857\,\rm{GHz}}< 1\,$Jy. We confined the analysis to the 1333 sources brighter than 1\,Jy. A cross-match with NED-LVS yields 1049 matches within 1\,arcmin. Conservatively restricting ourselves to distances $D<180\,$Mpc leaves 1022 sources. We obtain exactly the same number of matches using HECATE instead of NED-LVS. This suggests that both catalogs are complete for the local \textit{Planck}-detected galaxies, which are bright. However, the two samples are not identical: 28 sources differ. We expected a small difference because of the slightly different recipes used to compute the distances, implying that some sources moved below or above the chosen distance limit of 180 Mpc.

Here and for the other \textit{Planck} channels, we chose the cross-match with NED-LVS, considering the improved completeness of redshift and $z$-independent distance measurements. A close inspection of the sample identified five sources that are not galaxies and were therefore removed. They are PCCS2 857 G289.93+64.36 (blazar),  PCCS2E 857 G144.38-49.91 (cirrus), PCCS2 857 G208.54+33.29 (planetary nebula), PCCS2 857 G075.84-73.62 (HII region), and PCCS2E 857 G133.69-46.66 (molecular cloud). We kept the 32 sources classified as multiples (30 pairs, two triplets). If the far-infrared (FIR)/submm emission is not dominated by one member of the system,  the bright tail of the LF, where these sources reside, is slightly overestimated. Ignoring them would cause a slight incompleteness.

The surface density of NED-LVS galaxies within 180\,Mpc is $\simeq 5.9\,\hbox{deg}^{-2}$, implying that the expected number of random associations within $1'$ for 1017 trials is $\simeq 1.7$, i.e., $\simeq 0.17\%$. Extending the search radius to $2'$ adds 12 matches.  Using the Aladin sky atlas \citep{Bonnarel2000}, we checked whether these additional matches are real counterparts to \textit{Planck} sources. We find that two very extended, nearby galaxies (NGC\,55 and M81) are real counterparts; they were added to the sample, which then includes 1019 galaxies. The other matches are either Galactic sources or blazars, dominated by nonthermal radio emission from the active nucleus.

Figure\,\ref{LF857} shows our LLF estimates, obtained with the $1/V_{\rm max}$ and KDE methods. The results of the two methods are fully consistent with each other, with the central KDE estimates slightly higher at high luminosity. The figure also shows the estimates of \citet{Marchetti2016} and \citet{Negrello2013} for comparison. The latter are based on \textit{Planck}/ERCSC data, while the former rely on the \textit{Herschel} Multi-tiered Extragalactic Survey (HerMES) covering $39\,\hbox{deg}^2$, carried out with the SPIRE instrument.

We reach luminosities about a factor of three higher than \citet{Marchetti2016} and have statistical and systematic errors much lower than those of \citet{Negrello2013}. We find a significantly steeper decline above the knee luminosity than in earlier estimates.

We fitted the $1/V_{\rm max}$ results with the double power-law functions introduced by \citet{Lawrence1986}:
\begin{equation}
    \label{eq:Lawrence}
 \Phi(L)=\Phi_\star \left(\frac{L}{L_\star}\right)^{1-\alpha} \left(1+\frac{L}{\beta\, L_\star }\right)^{-\beta}\ \hbox{Mpc}^{-3}\,\hbox{dex}^{-1}\,,
   \end{equation}
and by \citet{Massardi2010}:
   \begin{equation}\label{eq:Massardi}
      \Phi(L)=\frac{\Phi_\star}{(L/L_\star)^\alpha+(L/L_\star)^\beta}\ \hbox{Mpc}^{-3}\,\hbox{dex}^{-1}\,.
   \end{equation}
Both fits are similarly good (reduced $\chi^2\simeq 0.4$). We also used Eq.\,(\ref{eq:Massardi}) to fit the KDE results. Table\,\ref{table:LFfit} lists the best-fit values of the parameters.

Table\,\ref{table:rho}. shows the total local luminosity densities obtained by integrating from zero to $\infty$ using eq.~(\ref{eq:Lawrence}), eq.~(\ref{eq:Massardi}), and the KDE LF. We note that the KDE value refers to the central estimates (solid dark blue line in Fig.\,\ref{LF857}), not to the double power-law fit (solid red line in the same figure); the latter gives a value that is 6\% higher. Since the low-luminosity slope is flat and the high-luminosity slope is steep, the luminosity density is only weakly sensitive to the integration limits. The integral from zero to infinity is only 3\% higher than the integral over the range covered by observations. Our luminosity density is $\simeq 35\%$ lower than the estimate of \citet{Marchetti2016}.

The exponential function of \citet{Saunders1990} gave a substantially worse fit and was not considered further.

\begin{table}
    \centering
    \caption{Estimates of the submm local luminosity densities, compared with those by \citet{Marchetti2016}.}
    \label{table:rho}
    \renewcommand{\arraystretch}{1}
\resizebox{\columnwidth}{!}{%
    \begin{tabular}{lcccc}
    \hline
    \hline
 & \multicolumn{1}{c}{$\log_{10}(\rho_{857\,\rm GHz})$} & \multicolumn{1}{c}{$\log_{10}(\rho_{545\,\rm GHz})$} & \multicolumn{1}{c}{$\log_{10}(\rho_{353\,\rm GHz})$} &
\multicolumn{1}{c}{$\log_{10}(\rho_{217\,\rm GHz})$} \\
 & $L_\odot\,\hbox{Mpc}^{-3}$   & $L_\odot\,\hbox{Mpc}^{-3}$  & $L_\odot\,\hbox{Mpc}^{-3}$ & $L_\odot\,\hbox{Mpc}^{-3}$   \\
    \hline
    Eq.\,(\ref{eq:Lawrence}) & 6.49 & 5.89 &  5.23 & 4.86  \\
    Eq.\,(\ref{eq:Massardi}) & 6.51 & 5.87 &  5.21 & 4.80 \\
    KDE                      & 6.53 & 5.94 &  5.24  & 4.77\\
    \hline
    Marchetti    & 6.64 & 5.96 & \\
   \hline
    \end{tabular}
    }
\tablefoot{The value at $500\,\mu$m given in the latter paper was converted to $550\,\mu$m (545 GHz), as explained in the text.}
\end{table}

\begin{figure}[ht!]
   \resizebox{\hsize}{!}
            {\includegraphics {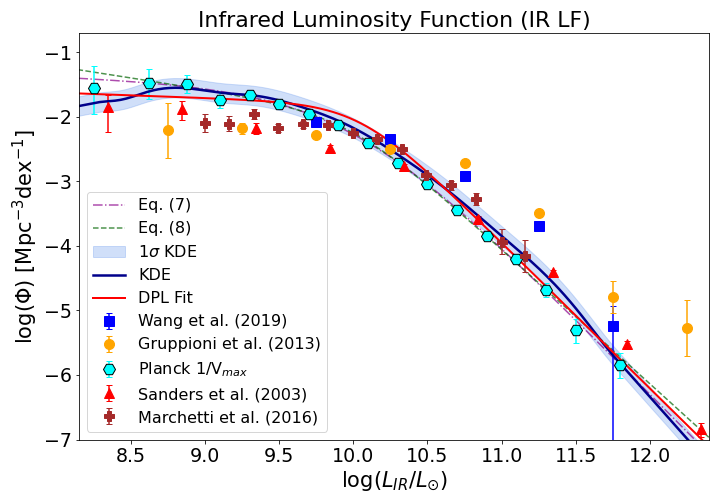}}
      \caption{Total IR (8--$1000\,\mu$m) local luminosity function determined using both the $1/V_{\rm max}$ and the KDE estimators. In the $1/V_{\rm max}$ case we also show the double power-law fits with eqs.\,(\ref{eq:Lawrence}) and (\ref{eq:Massardi}). In the KDE case, the solid dark blue line shows the central estimates, while the red line (DPL fit) shows the double power-law fit (eq.\,(\ref{eq:Massardi})). Earlier estimates by \citet{Sanders2003}, \citet{Gruppioni2013}, \citet{Marchetti2016}, and \citet{Wang2019}, scaled to the adopted value of $H_0$, are also shown for comparison. }
         \label{TIR_LF}
\end{figure}

\begin{table}[ht!]
\caption{Local IR (8--$1000\,\mu$m) LF of galaxies obtained with the $1/V_{\rm max}$ method.}                 
\label{table:IRLLF}    
\centering
\setlength{\tabcolsep}{5pt}                        
\begin{tabular}{c r r c r r}
\hline\hline               
\multicolumn{1}{c}{$\log(L_{\rm IR}$)} & \multicolumn{1}{c}{$N$} & \multicolumn{1}{c}{$N_{\rm eff}$} & \multicolumn{1}{c}{$\log(\Phi)$}  & \multicolumn{1}{c}{err$+$} & \multicolumn{1}{c}{err$-$} \\
\multicolumn{1}{c}{$L_\odot$} & & & \multicolumn{1}{c}{$\hbox{Mpc}^{-3}\,\hbox{dex}^{-1}$} \\
\hline                      
     8.250   &   3  &   2.355  &  -1.556  &   0.335  &   0.400\\
     8.625   &   7  &   5.291  &  -1.484  &   0.217  &   0.236\\
     8.875   &  17  &  11.679  &  -1.491  &   0.142  &   0.148\\
     9.100   &  20  &  14.816  &  -1.740  &   0.125  &   0.129\\
     9.300   &  47  &  39.275  &  -1.664  &   0.074  &   0.075\\
     9.500   &  72  &  59.181  &  -1.809  &   0.060  &   0.060\\
     9.700   &  92  &  74.884  &  -1.952  &   0.053  &   0.053\\
     9.900   & 125  &  99.236  &  -2.126  &   0.046  &   0.046\\
    10.100   & 142  & 124.447  &  -2.408  &   0.041  &   0.041\\
    10.300   & 127  & 102.524  &  -2.722  &   0.045  &   0.045\\
    10.500   & 116  &  91.961  &  -3.037  &   0.047  &   0.048\\
    10.700   &  92  &  71.333  &  -3.443  &   0.054  &   0.055\\
    10.900   &  64  &  50.771  &  -3.850  &   0.065  &   0.065\\
    11.100   &  51  &  38.949  &  -4.203  &   0.075  &   0.075\\
    11.300   &  26  &  20.670  &  -4.685  &   0.104  &   0.107\\
    11.500   &   9  &   7.530  &  -5.308  &   0.180  &   0.191\\
    11.800   &   7  &   7.000  &  -5.844  &   0.187  &   0.199\\
\hline                                  
\end{tabular}
\tablefoot{The columns have the same meaning as in Table\,\ref{table:857LLF}}
\end{table}

\begin{table}
    \centering
    \caption{Best-fit parameters for the double power-law fits to the total IR LLF.}
    \label{table:IRfit}
    \renewcommand{\arraystretch}{1}
    \begin{tabular}{lccrr}
    \hline
    \hline
Model & \multicolumn{1}{c}{$\log_{10}(\Phi_\star)$} & \multicolumn{1}{c}{$\log_{10}(L_\star)$} & \multicolumn{1}{c}{$\alpha$} & \multicolumn{1}{c}{$\beta$} \\
  & \multicolumn{1}{c}{$\hbox{Mpc}^{-3}\,\hbox{dex}^{-1}$} & \multicolumn{1}{c}{$L_\odot$} & &  \\
    \hline
    Eq.\,(\ref{eq:Lawrence}) & -1.536 & 9.626 & 1.098  & 2.233 \\
    Eq.\,(\ref{eq:Massardi}) & -1.928 & 9.982 &  0.358 &  2.085\\
    KDE                      & -1.796 & 10.046 & 0.083 & 2.259 \\
     \hline
    \end{tabular}
\end{table}

\subsection{The IR LLF}

We used the sample of 1019 galaxies selected at 857\,GHz to compute the total IR luminosity function.  Since the \textit{Planck} data do not encompass the dust emission peak of local galaxies, which generally occurs at $\sim 100 - 150\,\mu$m, we complemented them with photometry at shorter wavelengths to estimate the IR luminosity, $L_{\rm IR}$ (8--$1000\,\mu$m). At $100\,\mu$m we preferentially used the BeeP flux density derived from the IRIS map, which is missing for only 67 sources.

To better constrain the dust emission spectrum, we searched for IRAS $60\,\mu$m data. The cross-match with the IRAS Faint Source Catalog \citep[IRAS-FSC;][]{MoshirIRAS_FSC1990} yielded 947 matches within 2 arcmin. Only three of these are at a separation $>90^{\prime\prime}$; we checked that they are real associations. We find six of the 72 IRAS-FSC unmatched galaxies to have counterparts in the IRAS Small Scale Structure Catalog \citep{HelouWalker1988}. We retrieved IRAS flux densities from the NED at $60\,\mu$m for 12 additional sources and at $100\,\mu$m for seven more.

Overall, we find IRAS $100\,\mu$m flux densities for 31 of the 67 galaxies missing IRIS flux densities, so that only 36 lack $100\,\mu$m measurements; five of the latter have $160\,\mu$m photometry obtained by the Multiband Imaging Photometer for \textit{Spitzer} (MIPS) and one more has $170\,\mu$m photometry obtained by the Infrared Space Observatory (ISO). We retrieved IRAS $60\,\mu$m photometry for 968 galaxies; seven of the remaining 54 have $70\,\mu$m MIPS photometry. Only 30 galaxies (2.9\%) lack flux densities at or at shorter wavelengths than the dust emission peak.

A cross-match of the sample with the \textit{Herschel}/PACS Point Source Catalog \citep{Marton2024} yielded one match at $70\,\mu$m, 281 matches at $100\,\mu$m, and 170 matches at $160\,\mu$m. The $100\,\mu$m measurements allowed us to assess the effect of very different resolutions: the typical PACS full widths at half maximum (FWHMs) are $5.5''$, $6.7''$, and $11''$ in the 70, 100, and $160\,\mu$m bands, respectively. The median ratio between the IRIS and PACS $100\,\mu$m flux densities is $\simeq 2.1$ (the mean is 6, with a dispersion of 17.7). In this situation, we decided not to use PACS photometry. Specifically, out of the 30 galaxies lacking data at or shortward of the dust emission peak, only three have a PACS detection at 100 $\mu$m. Given the severe flux discrepancies and resolution issues discussed above, we chose to omit the PACS data even for these three sources. At the same wavelength, the mean ratio of IRIS to IRAS flux densities (angular resolution $\sim 2'$\footnote{see \url{https://lambda.gsfc.nasa.gov/product/iras/}}) is 0.95 with a dispersion of 0.22.

We computed the total IR luminosity by fitting the SED of starburst galaxies by \citet{Cai2013} to these data. The only parameter was normalization.

The small number of photometric points,  together with their rather large uncertainties, especially in the \textit{Planck} data, did not allow us to use more parameters for the fit. To overcome this limitation, we attempted to exploit also the optical/near-IR photometry, available for almost all our sources, to take advantage of codes interpreting the full SED while ensuring the balance between the energy absorbed and re-emitted by dust. This did not work: the two data sets turned out to be incompatible, and we could not find acceptable global fits respecting the energy balance. One reason for this is the strongly different resolutions between short- and long-wavelength observations. Our galaxies are quite extended, with sizes of up to several tens of arcminutes, although most of them are unresolved by \textit{Planck} (HFI beams have a FWHM of $\sim 5'$). In contrast, much higher optical/near-IR resolution requires large and highly uncertain aperture corrections, which frequently turn out to be insufficient (a situation similar to that encountered at $100\,\mu$m). Moreover, the colors in the latter bands indicate that in most cases measurements are dominated by regions with low dust obscuration; in other words, the optical/near-IR colors are not consistent with the amount of dust obscuration necessary to account for the submm measurements. \citet{Whittam2025} reached similar conclusions in a different framework.

Figure~\ref{TIR_LF} shows our estimates of the total IR luminosity function obtained with both the $1/V_{\rm max}$ and KDE methods. The results are listed in Table\,\ref{table:IRLLF}; we computed the $V_{\rm max}$ using the flux density at 857\,GHz, the \textit{Planck} frequency closest to the emission peak. Table\,\ref{table:IRfit} lists the best-fit values of the parameters. Again, both fits of eq.~(\ref{eq:Lawrence}) and eq.~(\ref{eq:Massardi}) are similarly good.

Our IR LLF spans a luminosity range similar to that of the estimate by \citet{Sanders2003}, derived from the IRAS Revised Bright Galaxy Sample, and is in reasonably good agreement with it. For $\log(L_{\rm IR}/L_\odot)\la 10$ the IR LLFs from \textit{Herschel} surveys are flatter than both the IRAS LLF and ours. At high luminosity, the latter LLFs move from above to below those of \citet{Gruppioni2013} and \citet{Wang2019}, while that of \citet{Marchetti2016} is intermediate between the IRAS LLF and ours.

The log of the local IR luminosity densities, $\log(\rho_{\rm IR}/L_\odot\,\hbox{Mpc}^{-3})$, obtained by fitting eq.~(\ref{eq:Lawrence}) and eq.~(\ref{eq:Massardi}), are 7.965 and 7.987, respectively. Our results agree well with the estimate by \citet{Marchetti2016} for the redshift range $0.02< z<0.1$, based on the HerMES survey over an area of $39\,\hbox{deg}^2$, $\log(\rho_{\rm IR}/L_\odot\,\hbox{Mpc}^{-3})=7.92$. \citet{Gruppioni2013} obtained a slightly higher value, $\log(\rho_{\rm IR}/L_\odot\,\hbox{Mpc}^{-3})=8.134$,for a \textit{Herschel}/PACS-selected sample with SPIRE data; however, this value refers to a wider redshift interval ($0< z<0.3$), where evolutionary effects start to be significant.

\begin{table}
\caption{Local luminosity function of galaxies at 545\,GHz ($550\,\mu$m) obtained with the $1/V_{\rm max}$ method.}                 
\label{table:545LLF}    
\centering
\setlength{\tabcolsep}{5pt}                       
\begin{tabular}{c r r c r r}
\hline\hline               
\multicolumn{1}{c}{$\log(L_{545\,\rm GHz}$)} & \multicolumn{1}{c}{$N$} & \multicolumn{1}{c}{$N_{\rm eff}$} & \multicolumn{1}{c}{$\log(\Phi)$}  & \multicolumn{1}{c}{err$+$} & \multicolumn{1}{c}{err$-$} \\
\multicolumn{1}{c}{$\hbox{W}\,\hbox{Hz}^{-1}$} & & & \multicolumn{1}{c}{$\hbox{Mpc}^{-3}\,\hbox{dex}^{-1}$} \\
\hline                      
    20.750    &   2  &   1.714  &  -0.959  &   0.394  &   0.501	\\
    21.250    &   4  &   3.702  &  -1.673  &   0.263  &   0.296	\\
    21.750    &  21  &  15.848  &  -1.666  &   0.120  &   0.124	\\	
    22.100    &  25  &  24.111  &  -1.654  &   0.096  &   0.098	\\	
    22.300    &  24  &  22.946  &  -1.986  &   0.099  &   0.101	\\	
    22.500    &  47  &  44.988  &  -1.967  &   0.069  &   0.070	\\	
    22.700    &  79  &  75.761  &  -2.035  &   0.053  &   0.053	\\	
    22.900    &  74  &  71.197  &  -2.379  &   0.054  &   0.055	\\	
    23.100    &  60  &  57.937  &  -2.779  &   0.060  &   0.061	\\	
    23.300    &  57  &  55.199  &  -3.080  &   0.062  &   0.063	\\	
    23.500    &  25  &  23.985  &  -3.758  &   0.096  &   0.098	\\	
    23.700    &  21  &  20.358  &  -4.210  &   0.105  &   0.108	\\	
    24.000    &   9  &   8.183  &  -5.223  &   0.172  &   0.182	\\
    24.350    &   5  &   4.996  &  -5.859  &   0.224  &   0.245	\\
\hline                                  
\end{tabular}
\end{table}

\begin{figure}
   \resizebox{\hsize}{!}
            {\includegraphics {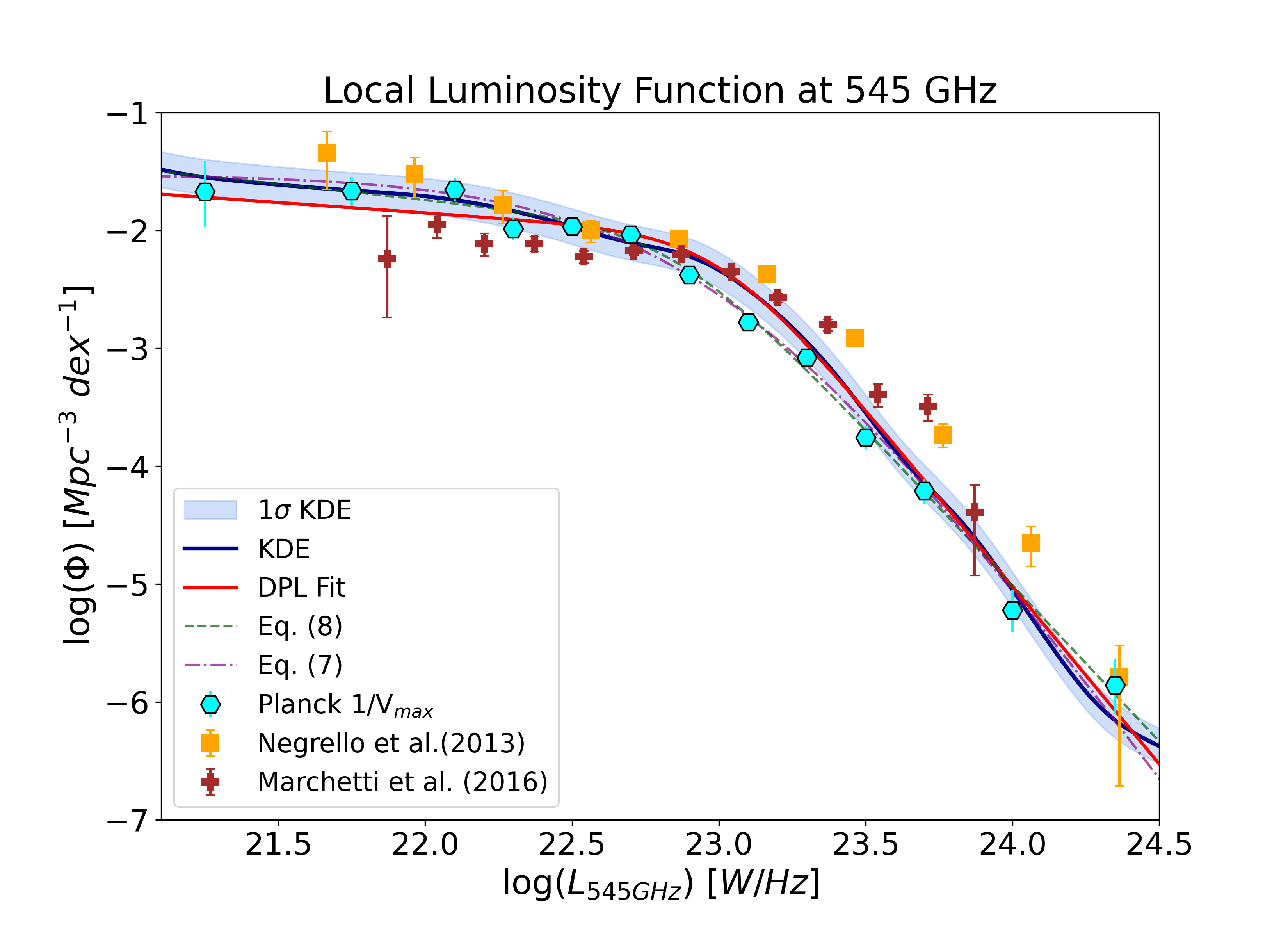}}
      \caption{Local luminosity function at 545\,GHz determined using both the $1/V_{\rm max}$ and KDE estimators, and double power-law fits, specified in the inset. The solid dark blue line shows the central KDE estimates, while the red line (DPL fit) shows the double power-law fit (eq.\,(\ref{eq:Massardi})). The earlier estimates by \citet{Negrello2013} and \citet{Marchetti2016}, scaled to the adopted value of $H_0$, are also shown for comparison. }
         \label{LF545}
\end{figure}

\subsection{The 545\,GHz ($550\,\mu$m) LLF}

The PCCS2 90\% completeness limit at 545 GHz is 555 mJy \citep[Table 1 of][]{PCCS22016}. For PCCS2E the limit is almost the same (557 mJy). We selected sources at $|b|>30^\circ$, $\hbox{APERFLUX} > 560\,$mJy and $\hbox{S/N}>5$. This initial sample contained 697 sources (540 from the PCCS2 and 157 from the PCCS2E), 484 of which  (443 from the PCCS2 plus 41 from the PCCS2E) have a NED-LVS counterpart within $2'$.

We restricted our sample to $D<180\,$Mpc and obtained 460 matches, 50 of which are at $>1'$ from the PCCS position. We find all the objects within $1'$ to be bona fide galaxies. We used the Aladin sky atlas to inspect the sources at $>1'$ individually. We find that one is a planetary nebula and six reside in regions heavily contaminated by cirrus emission. The NED photometric data confirmed that these sources would not have met our detection threshold without the cirrus contribution; we therefore removed them from the sample, leaving 453 galaxies. The higher fraction of matches at $>1'$ compared to 857 GHz may be due to the better positional accuracy of the BeeP catalog.

We computed the LLF as described above. We present the results in Table\,\ref{table:545LLF} and display them in Fig.\,\ref{LF545}, where we compare them with those of \citet{Negrello2007} and \citet{Marchetti2016}. The latter were computed from the HerMES survey at $500\,\mu$m. The luminosities at this wavelength were scaled down by a factor of 1.35 to convert them to 545\,GHz \citep{Maddox2018}.

We list the best-fit parameters for the fits of the $1/V_{\rm max}$ and KDE results in Table\,\ref{table:LFfit}. Around the knee luminosity, $L_\star$, the KDE estimate slightly exceeds the $1/V_{\rm max}$ estimate, perhaps due to smoothing by the bandwidth parameter.

Our results agree with earlier estimates by \citet{Negrello2013} and \citet{Marchetti2016} below $L_\star$, but at higher luminosities they are consistently lower. Similarly, we find a local luminosity density that is somewhat lower than that reported by \citet[][see Table\,\ref{table:rho}]{Marchetti2016}.

\begin{table}[ht!]
\caption{Local luminosity function of galaxies at 353\,GHz ($850\,\mu$m) obtained with the $1/V_{\rm max}$ method.}                 
\label{table:353LLF}    
\centering
\setlength{\tabcolsep}{5pt}                        
\begin{tabular}{c r r c r r}
\hline\hline               
\multicolumn{1}{c}{$\log(L_{353\,\rm GHz})$} & \multicolumn{1}{c}{$N$} & \multicolumn{1}{c}{$N_{\rm eff}$} & \multicolumn{1}{c}{$\log(\Phi)$}  & \multicolumn{1}{c}{err$+$} & \multicolumn{1}{c}{err$-$} \\
\multicolumn{1}{c}{$\hbox{W}\,\hbox{Hz}^{-1}$} & & & \multicolumn{1}{c}{$\hbox{Mpc}^{-3}\,\hbox{dex}^{-1}$} \\
\hline                      
    21.250   &      9  &    7.581  &   -1.647  &    0.179  &    0.190	\\
    21.625   &     15  &   14.642  &   -1.671  &    0.126  &    0.130	\\
    21.875   &     11  &   10.375  &   -2.143  &    0.151  &    0.158	\\
    22.100   &     32  &   30.652  &   -1.912  &    0.085  &    0.086	\\
    22.300   &     37  &   35.599  &   -2.131  &    0.078  &    0.079	\\
    22.500   &     38  &   36.578  &   -2.451  &    0.077  &    0.078	\\
    22.700   &     31  &   30.114  &   -2.804  &    0.085  &    0.087	\\
    22.900   &     12  &   11.817  &   -3.503  &    0.141  &    0.147	\\
    23.250   &      9  &    8.060  &   -4.439  &    0.173  &    0.183	\\
    24.000   &      4  &    3.094  &   -6.243  &    0.290  &    0.333	\\
\hline                                  
\end{tabular}
\end{table}

\begin{figure}[ht!]
   \resizebox{\hsize}{!}
            {\includegraphics {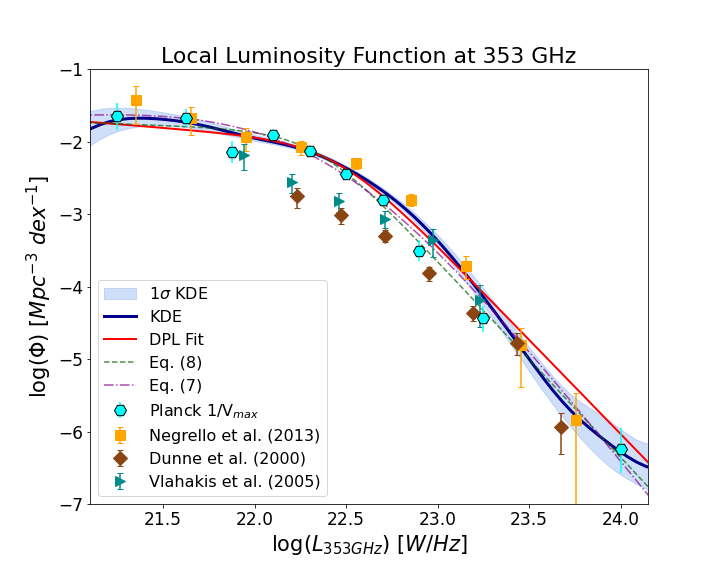}}
      \caption{Local luminosity function at 353\,GHz determined using both the $1/V_{\rm max}$ and KDE estimators, and double power-law fits, specified in the inset. The earlier estimates by \citet{Dunne2000} and \citet{Vlahakis2005}, both based on SLUGS data, and by \citet{Negrello2013}, all scaled to the adopted value of $H_0$, are also shown for comparison. }
         \label{LF353}
\end{figure}

\subsection{The 353\,GHz ($850\,\mu$m) LLF}

According to Table\,1 of \citet{PCCS22016}, the 90\% completeness limits at 353\,GHz are 304\,mJy for PCCS2 and 311\,mJy for PCCS2E. We find 888 (700 PCCS2 and 188 PCCS2E) sources with $\hbox{APERFLUX} > 320\,$mJy, $|b|> 30^\circ$ and $\hbox{S/N}>5$. The cross-match with NED-LVS yielded 280 matches within 2 arcmin (242 PCCS2 and 38 PCCS2E), of which 208 are within 1 arcmin. At distances $\le 180\,$Mpc we find 179 sources within 1 arcmin and 215 within 2 arcmin. We identified eight of the latter as being dominated by nuclear radio emission and removed them from the sample. We also removed two of the matches at distances $<1'$: one was not classified as a galaxy, and the other is too faint to be a credible counterpart to the \textit{Planck} source.

We inspected all 27 matches with separations larger than 1 arcmin in Aladin, and find that five are not valid counterparts to \textit{Planck} sources. We removed these five sources, leaving a sample of 200 galaxies.

We present our estimates of the LLF, computed as described above, in Fig.\,\ref{LF353} and in Table\,\ref{table:353LLF}.  Table\,\ref{table:LFfit} lists the parameters of the analytic fits and Table\,\ref{table:rho} gives the luminosity density.

Our results improve upon those of \citet{Negrello2013} and extend to luminosities about an order of magnitude fainter than the estimates by \citet{Dunne2000} and \citet{Vlahakis2005}, both based on the SCUBA Local Universe Galaxy Survey (SLUGS). There are hints of incompleteness of the SLUGS survey at $\log(L_{353\,\rm GHz}/\hbox{W}\,\hbox{Hz}^{-1})\le 22.5$.

\begin{table}[ht!]
\caption{Local dust mass function derived from the 353\,GHz ($850\,\mu$m) LLF.}                 
\label{table:LDMF}    
\centering
\setlength{\tabcolsep}{5pt}                        
\begin{tabular}{c c r r}
\hline\hline               
\multicolumn{1}{c}{$\log(M_{\rm dust}$)} & \multicolumn{1}{c}{$\log(\Phi_{\rm dust})$} &  \multicolumn{1}{c}{err$+$} & \multicolumn{1}{c}{err$-$} \\
\multicolumn{1}{c}{$M_\odot$} & \multicolumn{1}{c}{$\hbox{Mpc}^{-3}\,\hbox{dex}^{-1}$} \\
\hline                      
          6.359   &      -1.647  &    0.190  &    0.179	  \\
          6.734   &      -1.671  &    0.130  &    0.126	  \\
          6.984   &      -2.143  &    0.158  &    0.151	  \\
          7.209   &      -1.912  &    0.086  &    0.085	  \\
          7.409   &      -2.131  &    0.079  &    0.078	  \\
          7.609   &      -2.451  &    0.078  &    0.077	  \\
          7.809   &      -2.804  &    0.087  &    0.085	  \\
          8.009   &      -3.503  &    0.147  &    0.141	  \\
          8.359   &      -4.439  &    0.183  &    0.173	  \\
          9.109   &      -6.243  &    0.333  &    0.290	  \\
\hline                                  
\end{tabular}
\tablefoot{Derived using $T_{\rm d}= 17.7\,$K and $\kappa_{850\mu\rm m}=0.77$ (see eq.\,(\ref{eq:dust_mass})).}
\end{table}

\begin{table}
    \centering
    \caption{Best-fit parameters for the double power-law fits to the dust mass function. }
    \label{table:DMfit}
    \renewcommand{\arraystretch}{1}
    \begin{tabular}{lccrr}
    \hline
    \hline
Model & \multicolumn{1}{c}{$\log_{10}(\Phi_\star)$} & \multicolumn{1}{c}{$\log_{10}(M_\star)$} & \multicolumn{1}{c}{$\alpha$} & \multicolumn{1}{c}{$\beta$} \\
  & \multicolumn{1}{c}{$\hbox{Mpc}^{-3}\,\hbox{dex}^{-1}$} & \multicolumn{1}{c}{$M_\odot$} & &  \\
    \hline
    Eq.\,(\ref{eq:Lawrence}) & -1.400 & 6.973 & 0.790 & 3.437 \\
    Eq.\,(\ref{eq:Massardi}) & -1.854 & 7.445 &  0.093 & 2.705 \\
     \hline
    \end{tabular}
\end{table}

\begin{figure}[ht!]
   \resizebox{\hsize}{!}
            {\includegraphics {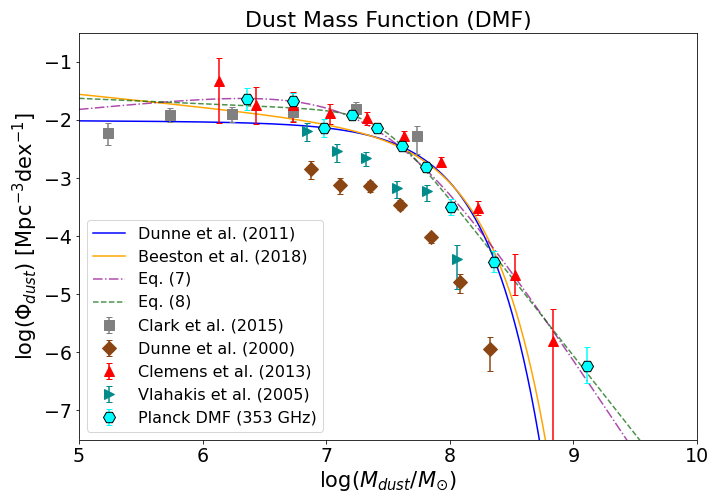}}
      \caption{Local dust mass function, with double power-law fits specified in the inset. Also shown for comparison are the estimates by \citet{Dunne2000}, \citet{Vlahakis2005}, \citet{Clemens2013}, and \citet{Clark2015}, as well as the best-fitting functions calculated by \citet{Dunne2011} and \citet{Beeston2018}. Previous results were scaled, when necessary, to the values of $H_0$ and $\kappa_{850\mu\rm m}$ adopted in this paper. }
         \label{fig:DMF}
\end{figure}.

\subsection{The dust mass function}

We computed the dust mass from the 353\,GHz ($850\,\mu$m) luminosity using the formula \citep{Hildebrand1983}
\begin{equation}\label{eq:dust_mass}
L_{850\mu\rm m}(T_{\rm d})=4\pi M_{\rm d}(T_{\rm d}) \kappa_{850\mu\rm m} B_\lambda(T_{\rm d}),
\end{equation}
where $T_{\rm d}$ is the dust temperature, $\kappa_{850\mu\rm m}$ is the dust absorption coefficient, and $B_\lambda(T_{\rm d})$ is the black-body function. Following \citet{Clemens2013} we adopted $T_{\rm d}=17.7\,$K. The value of $\kappa_{850\mu\rm m}$ is still debated. We chose $\kappa_{850\mu\rm m}=0.77\,\hbox{cm}^2\,\hbox{g}^{-1}$, following \citet{Dunne2000}, \citet{Vlahakis2005}, and \citet{Beeston2018}.

We then derived the local dust mass function (DMF) from the 353\,GHz ($850\,\mu$m) LLF. We present the results in Table\,\ref{table:LDMF} and Fig.\,\ref{fig:DMF}.  Table\,\ref{table:DMfit} shows the best-fit values of the double power-law parameters (eqs.\,(\ref{eq:Lawrence}) and (\ref{eq:Massardi})).

Figure\,\ref{fig:DMF} shows the DMF data points derived by \citet{Dunne2000}, \citet{Vlahakis2005}, \citet{Clemens2013}, and \citet{Clark2015} for comparison. In addition, we plotted the best-fitting Schechter functions obtained by \citet{Dunne2011} and \citet{Beeston2018}, whose data points are not tabulated. We scaled all estimates to the values of the parameters used here.

\citet{Dunne2000} and \citet{Vlahakis2005} used SLUGS observations of IRAS-selected or optically-selected samples, respectively. Both papers cautioned that these selections could miss a fraction of dusty galaxies; subsequent estimates based on blind surveys confirmed this concern.

\citet{Clemens2013} derived the dust masses using the public code Multiwavelength Analysis of Galaxy Physical Properties \citep[MAGPHYS;][]{daCunha2008}. To account for the different dust temperatures of the sources, they computed the bivariate $545\,$GHz--luminosity/dust mass function. They then obtained the local dust mass function from the LLF at $545\,$GHz built by \citet{Negrello2013} using the \textit{Planck} ERCSC, summing the contributions of all $545\,$GHz luminosities to each dust mass bin.

\citet{Clark2015} built a  sample of 42 nearby galaxies (distance in the range $15<D<46\,$Mpc), extracted from the \textit{Herschel} Astrophysical Terahertz
Large Area Survey \citep[H-ATLAS;][]{Eales2010} Phase 1 catalog \citep{Valiante2016}. They derived distances from spectroscopic redshifts, correcting for bulk deviations from the Hubble flow. To estimate dust temperatures and masses they fitted the \textit{Herschel} PACS and SPIRE plus the IRAS $60\,\mu$m photometry with two dust components, both represented by modified black bodies. They derived the dust mass function with the $1/V_{\rm max}$ method.

\citet{Beeston2018} computed the local dust mass function with the $1/V_{\rm max}$ method, applied to a sample of 15,750 galaxies in the redshift range $0.002<z<0.1$, drawn from the H-ATLAS survey. They determined distances as in \citet{Clark2015} and obtained dust masses using the MAGPHYS package.

The DMF by \citet{Beeston2018} spans the broadest range of dust masses, extending down to $\log(M_{\rm dust}/M_\odot)\sim 4$. Figure \,\ref{fig:DMF} shows that our low-mass slope is almost identical to theirs, although our normalization is slightly higher. We differ at the highest masses where the decline of our DMF is consistent with a power-law instead of an exponential, in agreement with the findings of \citet{Clemens2013}. Integrating our DMF, we obtain a local dust mass density of $\log(\rho_{\rm dust}/L_\odot\,\hbox{Mpc}^{-3})=5.38$ for the fit with eq.\,(\ref{eq:Lawrence}) and 5.36 with eq.\,(\ref{eq:Massardi}). The corresponding densities in units of the critical density are $\log(\Omega_{\rm dust})=-5.72$ or $-5.74$, respectively, which are somewhat higher than the value $\log(\Omega_{\rm dust})=-5.97$, found by \citet{Beeston2018}.

\begin{table}[ht!]
\caption{Local luminosity function of galaxies at 217\,GHz (1.38\,mm) obtained with the $1/V_{\rm max}$ method.}                 
\label{table:217LLF}    
\centering
\setlength{\tabcolsep}{5pt}                        
\begin{tabular}{c r r c r r}
\hline\hline               
\multicolumn{1}{c}{$\log(L_{217\,\rm GHz})$} & \multicolumn{1}{c}{$N$} & \multicolumn{1}{c}{$N_{\rm eff}$} & \multicolumn{1}{c}{$\log(\Phi)$}  & \multicolumn{1}{c}{err$+$} & \multicolumn{1}{c}{err$-$} \\
\multicolumn{1}{c}{$\hbox{W}\,\hbox{Hz}^{-1}$} & & & \multicolumn{1}{c}{$\hbox{Mpc}^{-3}\,\hbox{dex}^{-1}$} \\
\hline                      
    21.000    &     5  &   4.304  &  -1.510  &   0.243  &   0.269 \\
    21.500    &     8  &   6.676  &  -2.224  &   0.192  &   0.205 \\
    21.875    &    13  &  11.947  &  -2.131  &   0.140  &   0.146 \\
    22.250    &     7  &   5.920  &  -3.225  &   0.205  &   0.221 \\
\hline                                  
\end{tabular}
\end{table}

\begin{figure}[ht!]
   \resizebox{\hsize}{!}
            {\includegraphics {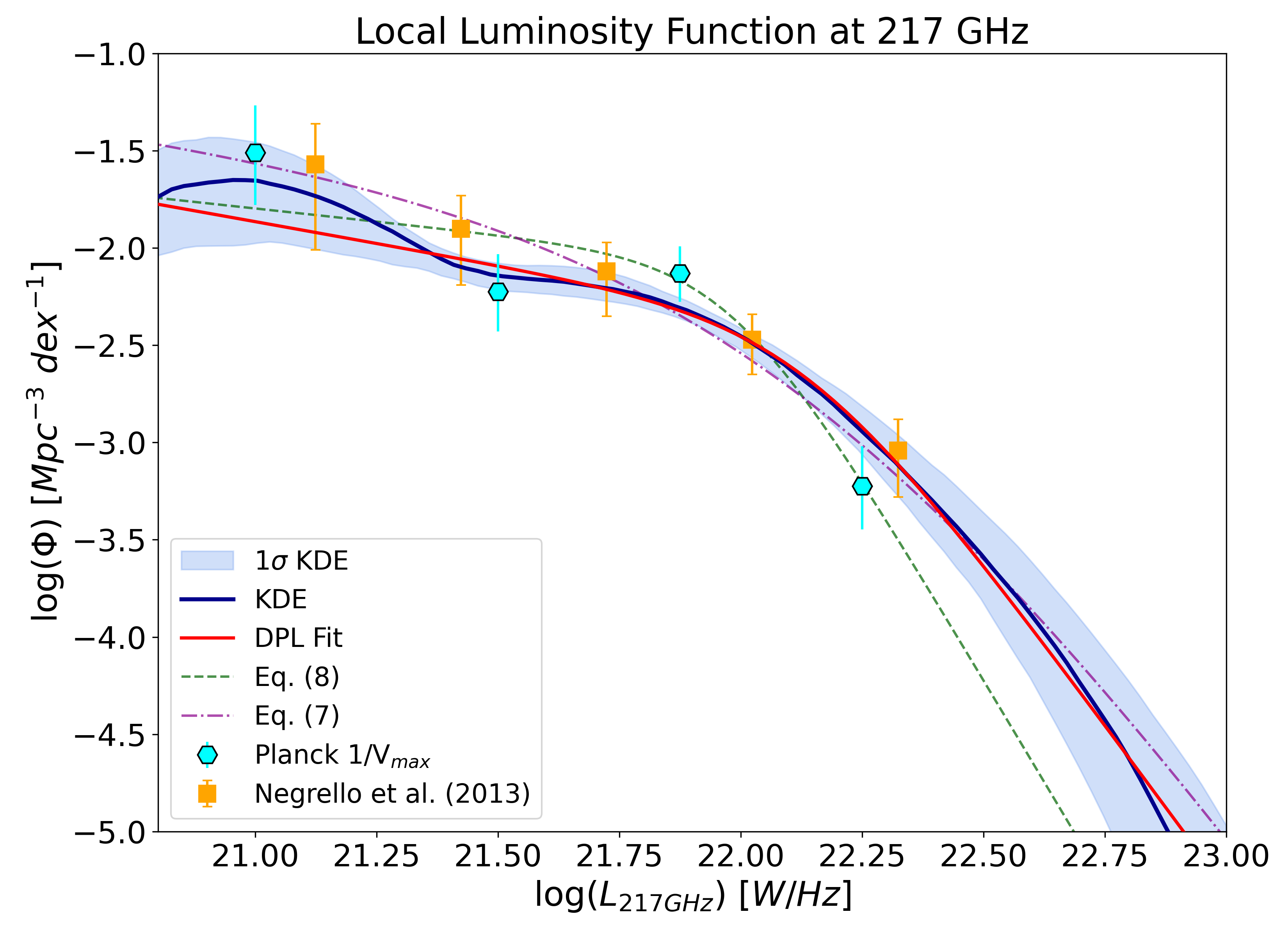}}
      \caption{Local luminosity function at 217\,GHz determined using both the $1/V_{\rm max}$ and the KDE estimators, and double power-law fits, specified in the inset. The earlier estimate by \citet{Negrello2013}, scaled to the adopted value of $H_0$, is also shown for comparison. }
         \label{LF217}
\end{figure}

\subsection{The 217\,GHz (1.38\,mm) LLF}

At this frequency, the 90\% PCCS2 completeness limit is 152\,mJy, while the PCCSE limit is unclear \citep{PCCS22016}. There are 235 PCCS2 and 72 PCCS2E sources at $|b|> 30^\circ$ and $\hbox{S/N}>5$; all have $\hbox{APERFLUX} > 250\,$mJy. We adopted this value as the flux limit.

The cross-match with NED-LVS (2\,arcmin radius) yields 42 matches within 180\,Mpc for the PCCS2 sample and an additional match for the PCCS2E sample. We removed PCCS2\,217\,G144.03+38.39 because its flux density is heavily contaminated by cirrus. We also removed the ten sources that, after a search of NED, were found to be dominated by nuclear radio emission. Using the interacting Aladin sky atlas, we verified the associations of the remaining 33 sources with NED-LVS galaxies, including the six sources with nominal separations larger than $1'$ (maximum separation $96''$). These six sources are all very extended galaxies, with sizes larger than $10'$.

 We present the derived LLF in Table\,\ref{table:217LLF} and Fig.\,\ref{LF217}. This figure shows that our results are in good agreement with the estimate of \citet{Negrello2013}. The best-fit values of the parameters of the two power-law fits are listed in Table\,\ref{table:LFfit}, and the corresponding local luminosity densities are given in Table\,\ref{table:rho}.

\section{Discussion and conclusions}\label{sect:conclusions}

The shallow \textit{Planck} surveys with the HFI, with their all-sky coverage (except for a zone above and below the Galactic plane) allow us to sample the truly local (sub)mm LF up to the rarest highest luminosities, being immune to the dramatic evolutionary effects present in this spectral region. These effects are already significant at $z\simeq 0.1$ \citep{Dunne2011, Marchetti2016}, that is, over the redshift range spanned by some \textit{Herschel} samples used to derive the LLF. Moreover, thanks to the large area, sampling variance due to inhomogeneities in the spatial distribution is reduced. Moreover, \textit{Planck} surveys extend to frequencies $< 600\,$GHz, where no other large-area survey exists. However, much deeper \textit{Herschel} surveys can, in principle, reach fainter luminosities.

Compared with the preliminary estimates of \cite{Negrello2013}, which were based on  \textit{Planck} ERCSC, our analysis benefits from substantially improved data. The use of full-mission maps (five HFI surveys instead of the 1.6 ERCSC surveys) and improved data processing increased the sensitivity (measured as the flux density of 90\% completeness) by almost a factor of three and the corresponding number of detections by almost an order of magnitude. The reliability of the detections is also better characterized.

Equally important, the all-sky catalogs of local galaxies, HECATE and NED-LVS, have provided a powerful tool for the optical identification of \textit{Planck} sources, providing redshifts for all galaxies and redshift-independent distances, whenever possible, for the nearest objects for which redshifts are not reliable distance indicators.

We computed the LLFs at 857, 545, 353, and 217\,GHz (350, 550, 850, and $1380\,\mu$m, respectively) with both the classical $1/V_{\rm max}$ and the nonparametric KDE method. The results from both approaches are in good agreement across the entire luminosity range. This consistency confirms the robustness of the derived spatial densities and demonstrates that the empirical optimization of the KDE bandwidth parameter described in Sect.\,\ref{SectLF} successfully suppresses Poisson noise without introducing systematic smoothing biases.

Below the knee luminosity of the LLF, our results agree with those derived from \textit{Herschel} surveys at 857 and 600\,GHz (350 and $500\,\mu$m) and with those of \citet{Negrello2013}. At higher luminosities, we find somewhat lower space densities. For \citet{Negrello2013}, this discrepancy is likely caused by boosting of ERCSC flux densities below $S_{857\,\rm GHz}\simeq 1.3$ by the \citet{Eddington1913} bias \citep{PlanckERCSC_Expl_Suppl2012, Herranz2013}. The origin of the discrepancy with \textit{Herschel}-based results is unclear. We note, however, that the LFs at 857 and 600\,GHz by \citet{Marchetti2016} were derived from a sample selected at $250\,\mu$m, with the luminosities at the other frequencies inferred by SED fitting, whereas our estimates rely on samples directly selected at each specific frequency. Furthermore, the \citet{Marchetti2016} sample covers a wider redshift range ($z \le 0.1$), where mild evolutionary effects might already begin to emerge \citep[see also][]{Dunne2011}, while our study is strictly restricted to the immediate local volume ($D < 180$\,Mpc). The multiwavelength selection framework adopted in that work, combining an $r$-magnitude limit with flux cuts at $24\,\mu$m (MIPS) and $250\,\mu$m (\textit{Herschel}), introduces additional selection effects that must be carefully modeled in the $V_{\rm max}$ evaluation. These combined factors introduce systematic uncertainties that are difficult to quantify and may not be fully captured by the nominal statistical error bars, making a rigorous comparison of the significance of the discrepancies between the two sets of estimates challenging.

The total IR LLF, derived from the LLF at 857\,GHz, is reasonably consistent with that derived from IRAS data \citep{Sanders2003}, which span a similar luminosity range. Interestingly, both estimates show a power-law decline at high luminosity, at variance with the steeper exponential (Schechter) decline generally assumed to fit \textit{Herschel}-derived LFs.

The IR LFs of \citet{Gruppioni2013} and \citet{Wang2019}, derived from \textit{Herschel} data, are substantially higher than ours and the IRAS-based LF above $\log(L_{\rm IR}/L_\odot)=10.5$. This is expected since their samples span the range $z=0$--0.3 and show strong evolution, already appreciable at $z=0.1$. Moreover, \citet{Gruppioni2013} adopted a $3\,\sigma$ detection limit, which makes the flux densities susceptible to flux boosting effects (\citealt{Eddington1913} bias). Below $\log(L_{\rm IR}/L_\odot)\simeq 9.5$, our results are above those of \citet{Marchetti2016} and \citet{Gruppioni2013}; again, the significance of the difference is difficult to assess, given the issues mentioned above.

At 353\,GHz we are again consistent with \citet{Negrello2013} up to the knee luminosity, while their LLF by is somewhat higher over the following decade in luminosity. Once again, the discrepancy can be explained by the boosting of ERCSC flux densities. Above $\log(L_{353\,\rm GHz}/\hbox{W}\,\hbox{Hz}^{-1})\simeq 23$ our results agree with those of \citet{Dunne2000} and \citet{Vlahakis2005}, based on the pointed SLUGS observations, which show indications of incompleteness at lower luminosity. The SLUGS samples, which were IRAS or optically selected, may easily miss faint 353\,GHz sources. In fact, the dust mass functions of \citet{Dunne2000} and \citet{Vlahakis2005}, derived from their 353\,GHz LF, are also well below those obtained by \citet{Dunne2011}, \citet{Clark2015}, and \citet{Beeston2018} from H-ATLAS samples. The latter estimates are in reasonably good agreement with ours, which are also derived from the 353\,GHz LF. The Schechter fits of the DMF by \citet{Dunne2011} and \citet{Beeston2018} deviate from our fits at the highest dust masses, but their samples do not sample the DMF above a few times $10^8\,M_\odot$. At 217\,GHz our results are in excellent agreement with \citet{Negrello2013}.

\begin{acknowledgements}
We are grateful to the anonymous referee for the accurate reading of the manuscript and many useful comments. M.B. acknowledges support from INAF under the mini-grant ``A systematic search for ultra-bright high-z strongly lensed galaxies in Planck catalogues''. This paper is based on observations obtained with \textit{Planck}\footnote{http://www.esa.int/Planck}, an ESA science mission with instruments and contributions directly funded by ESA Member States, NASA, and Canada.
We have made use of the TOPCAT package \citep{Taylor2005}, of the ``Aladin sky atlas'' developed at Centre de Donn\'ees astronomiques de Strasbourg (CDS), Strasbourg Observatory, France \citep{Bonnarel2000} and of the NASA/IPAC Extragalactic Database, which is funded by the National Aeronautics and Space Administration and operated by the California Institute of Technology.
\end{acknowledgements}

   \bibliographystyle{aa} 
   \bibliography{PlanckLLF.bib} 

\end{document}